\documentclass[amsmath,amssymb,floatfix,lengthcheck,nopreprintnumbers,prl,showpacs,superscriptaddress,twocolumn]{revtex4}
\usepackage{graphicx}

\newcommand{\beq}{\begin{equation}}
\newcommand{\eeq}{\end{equation}}
\newcommand{\beqs}{\begin{eqnarray}}
\newcommand{\eeqs}{\end{eqnarray}}

\begin{document}

\title{Parity Doubling and the S Parameter Below the Conformal Window}

\author{T.~Appelquist}
\affiliation{Department of Physics, Sloane Laboratory, Yale University,
             New Haven, Connecticut 06520, USA}
\author{R.~Babich}
\affiliation{Department of Physics, Boston University,
	Boston, Massachusetts 02215, USA}
\author{R.~C.~Brower}
\affiliation{Department of Physics, Boston University,
	Boston, Massachusetts 02215, USA}
\author{M.~Cheng}
\affiliation{Physical Sciences Directorate, Lawrence Livermore National Laboratory,
	Livermore, California 94550, USA}
\author{M.~A.~Clark}
\affiliation{Harvard-Smithsonian Center for Astrophysics, Cambridge, Massachusetts 02138, USA}
\author{S.~D.~Cohen}
\affiliation{Department of Physics, Boston University,
	Boston, Massachusetts 02215, USA}
\author{G.~T.~Fleming}
\affiliation{Department of Physics, Sloane Laboratory, Yale University,
             New Haven, Connecticut 06520, USA}
\author{J.~Kiskis}
\affiliation{Department of Physics, University of California,
	Davis, California 95616, USA}
\author{M.~F.~Lin}
\affiliation{Department of Physics, Sloane Laboratory, Yale University,
             New Haven, Connecticut 06520, USA}
\author{E.~T.~Neil}
\affiliation{Department of Physics, Sloane Laboratory, Yale University,
             New Haven, Connecticut 06520, USA}
\author{J.~C.~Osborn}
\affiliation{Argonne Leadership Computing Facility,
	Argonne, Illinois 60439, USA}
\author{C.~Rebbi}
\affiliation{Department of Physics, Boston University,
	Boston, Massachusetts 02215, USA}
\author{D.~Schaich}
\affiliation{Department of Physics, Boston University,
	Boston, Massachusetts 02215, USA}
\author{P.~Vranas}
\affiliation{Physical Sciences Directorate, Lawrence Livermore National Laboratory,
	Livermore, California 94550, USA}
\collaboration{Lattice Strong Dynamics (LSD) Collaboration}
\noaffiliation

\begin{abstract}
We describe a lattice simulation of the masses and decay constants of the lowest-lying vector and axial resonances, and the electroweak $S$ parameter, in an $SU(3)$ gauge theory with $N_f = 2$ and $6$ fermions in the fundamental representation. The spectrum becomes more parity doubled and the $S$ parameter per electroweak doublet decreases when $N_f$ is increased from $2$ to $6$, motivating study of these trends as $N_f$ is increased further, toward the critical value for transition from confinement to infrared conformality.
\end{abstract}


\maketitle

\paragraph{\textbf{Introduction}}
In a recent letter \cite{Appelquist:2009ka}, we studied the chiral properties of an SU$(3)$ gauge theory with $N_f$ massless Dirac fermions in the fundamental representation as $N_f$ is increased from 2 to 6. We noted that the $N_f = 2$ simulations are in good agreement with measured QCD values, and that the $N_f = 6$ results indicate substantial enhancement of the chiral condensate. Here we extend our study of these two theories, presenting results for the electroweak $S$ parameter and for the lightest vector and axial resonances.

There is evidence from lattice simulations \cite{Appelquist:2007hu, Fodor:2009wk, Nagai:2009ip, DelDebbio:2010hx, Hasenfratz:2010fi} that for an SU$(N)$ gauge theory, infrared conformality exists for a range of $N_f$ values from the onset of asymptotic freedom down to a critical value $N_f^c$. Below this ``conformal window", chiral symmetry breaking and confinement set in. Even for $N_f < N_f^c$ there can remain an approximate infrared fixed point provided that $0 < N_f^c - N_f \ll N_f^c$. The scale of chiral symmetry breaking is then small, and the fixed point approximately governs the theory from the breaking scale out to some higher scale. This ``walking" behavior leads to chiral-condensate enhancement, which can address the problem of obtaining large enough quark and lepton masses in technicolor theories.

It has been suggested \cite{Appelquist:1998xf, Hsu:1998jd, Kurachi:2006mu} that walking theories could address another problem by leading to smaller values of the electroweak $S$ parameter. The value of $S$ is related to the spectrum of vector and axial resonances in the theory.  As in Ref. \cite{Appelquist:2009ka}, we start with $N_f = 2$, allowing us to check the reliability of our methods by comparison with QCD phenomenology. Proceeding carefully toward $N_f^c$ is prudent since the emergence of widely separated scales associated with walking is challenging for lattice methods.

We first compute the $S$ parameter from the defining current correlators, and then present results for the lowest lying vector and axial masses and decay constants. We discuss our results along with the related Weinberg spectral function sum rules, and then summarize.

\paragraph{\textbf{The S parameter}}

The $S$ parameter can be defined in terms of the vector and axial current-correlation functions with, by convention, the would-be Nambu-Goldstone-boson (NGB) contribution to the standard-model (SM) radiative corrections removed. With $N_f/2$ massless electroweak doublets, it can be written as \cite{Peskin:1991sw}
\begin{align}
S &= 4\pi (N_f/2)\left[ \Pi'_{VV}(0) - \Pi'_{AA}(0) \right] - \Delta S_{SM}
\nonumber \\
&= \frac{1}{3\pi} \int_0^\infty \frac{ds}{s} \Bigg\{(N_f/2) \left[ R_V(s) - R_A(s) \right] \nonumber \\
&\hspace{5mm}\left.- \frac{1}{4} \left[ 1 - \left(1 - \frac{m_H^2}{s}\right)^3 \theta (s - m_H^2) \right] \right\}\,, \label{eq:Sdef}
\end{align}
where $\Pi_{VV}(Q^2)$ and $\Pi_{AA}(Q^2)$ are the transverse correlation functions
for a single electroweak doublet, $R(s) \equiv 12\pi$ Im $ \Pi'(s)$, and $m_H$ is the reference Higgs mass. The presence of $R_V(s) - R_A(s)$ in the spectral integral suggests that $S$ could decrease if the resonance spectrum becomes more parity doubled with increasing $N_f$.

For $N_f=2$, there are $3$ Goldstone bosons, with the $I_3=1$ pair leading to $R_V(s) \rightarrow 1/4$ as $s \rightarrow 0$. ($R_A(s) \rightarrow 0$.) The standard-model subtraction removes the resultant infrared divergence. With $N_f/2$ electroweak doublets, there are $N_f^2 -1 $ Goldstone bosons in the absence of other interactions. Among these, $(N_f/2)^2$ pairs contribute to $S$, leading to
$R_V(0) = (1/4)N_f/2$. With standard-model and other interactions included, $N_f^2 - 4$ of the $N_f^2 -1$ Goldstone bosons will be pseudo-Goldstone bosons (PNGBs). The $S$ parameter is again infrared finite, depending logarithmically on the masses of the PNGBs.

Lattice simulations are carried out with a finite fermion mass $m_f$, requiring extrapolation to reach the chiral limit. With $N_{f}/2$ electroweak doublets, since we do not include SM and other interactions to give mass to the PNGBs, the extrapolation for $ N_f \neq 2$ would lead to log $m_f$ terms in $S$. For our $N_f = 6$ simulations, $m_f$ is not yet small enough to see clear evidence for these chiral logs. For smaller $m_f$, the log $m_f$ terms would be replaced by logarithmic dependence on the PNGB masses in the full theory.

\paragraph{\textbf{Simulation Details}}

Simulations are performed using domain-wall fermions and the Iwasaki improved gauge action \cite{Allton:2008pn}.  The domain-wall formulation suppresses the chiral symmetry breaking associated with fermion discretization, and preserves flavor symmetry at finite lattice spacing, both desirable properties for computation of the $S$-parameter.  Gauge configurations are generated as in Ref. \cite{Appelquist:2009ka}.  Dimensionful quantities are given in lattice units.

The lattice volume is set to $32^3 \times 64$, with the length of the fifth dimension $L_s = 16$ and the domain-wall height $m_0 = 1.8$. The choices $\beta=2.70$ for $N_f = 2$ and $\beta= 2.10$ for $N_f = 6$ lead
to nearly the same physical scale in lattice units. Simulations are performed for fermion masses $m_f = 0.005$ to $0.03$, although the $N_f = 2$ results for $m_f = 0.005$ may suffer from finite-volume effects, and are not included in the analysis. At finite lattice spacing, even with $m_f = 0$, the chiral symmetry is not exact, with the violation captured in a residual mass $m_{res}\ll m_f$. The total fermion mass $m$ is then  $ m \equiv m_f + m_{res}$.

\paragraph{\textbf{Current Correlators}}

The lattice expression for the current correlator of interest is

\begin{eqnarray}
\Pi^{\mu\nu}_{VV}(Q) &=& \delta^{\mu \nu}\Pi_{VV}(Q^2) - (Q^{\mu} Q^\nu / Q^2)
{\widetilde\Pi}_{VV}(Q^2)
\nonumber \\
&=& Z \sum_x e^{iQ\cdot(x +
\hat\mu / 2)} \langle \mathcal V^{\mu}(x) V^{\nu}(0) \rangle
\end{eqnarray}
and similarly for $\Pi_{AA}$.  Here $\mathcal V^{\mu}$ is the conserved domain-wall vector current, $V^{\nu}$ is the non-conserved local current, and $Z$ is a non-perturbative renormalization constant.  $(x + \hat\mu / 2)$ appears because $\mathcal
V^{\mu}(x)$ is point split on the link $(x,x+\mu)$. The use of conserved currents ensures that lattice
artifacts cancel in the $V-A$ current correlator
$\Pi_{V-A}(Q^2) \equiv \Pi_{VV}(Q^2) - \Pi_{AA}(Q^2)$ \cite{Boyle:2009xi}.

\begin{figure}
  \includegraphics[width=85mm]{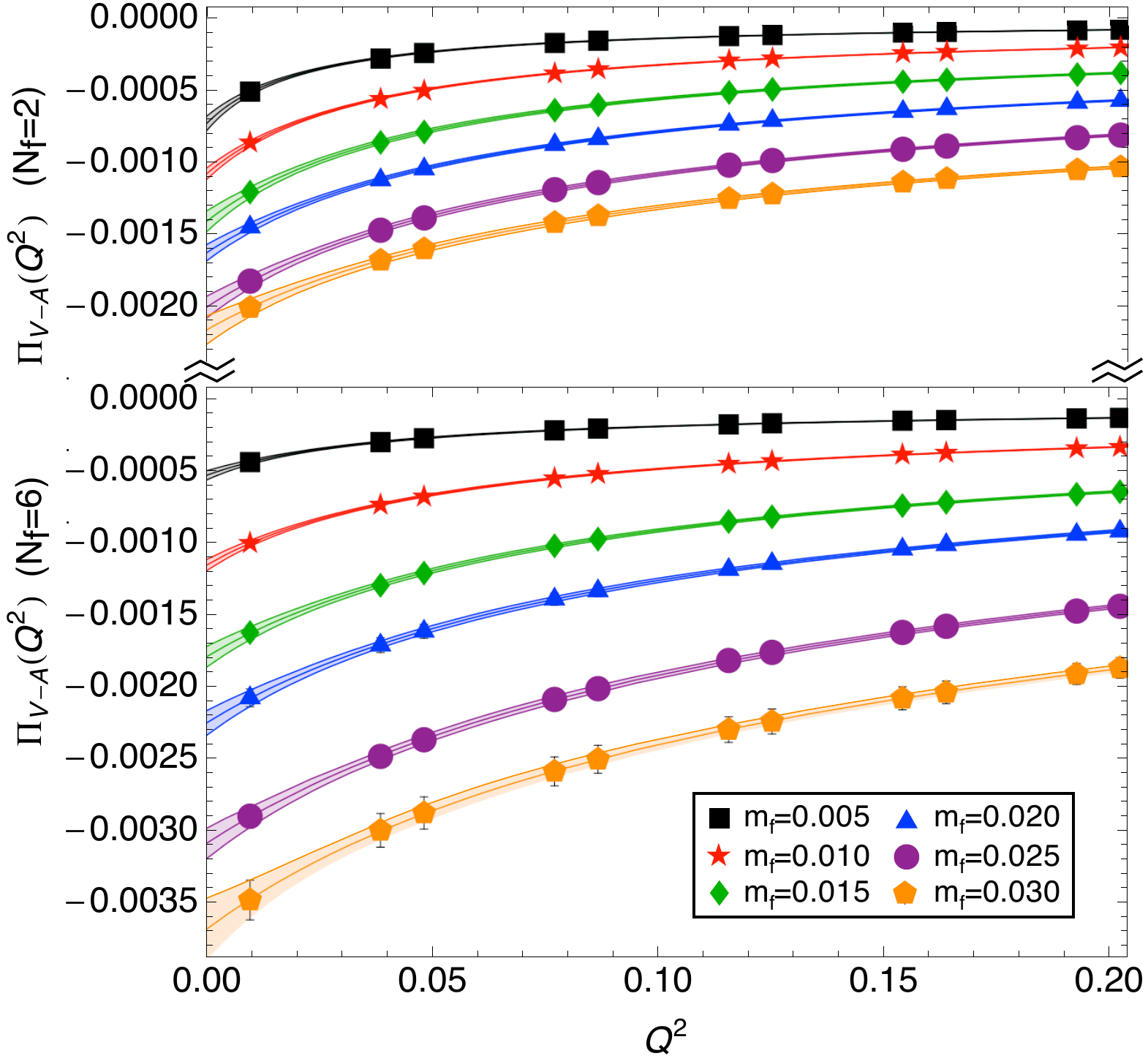}
  \caption{\label{fig:V-A}$\Pi_{V-A}(Q^2)$ data and fits for $N_f = 2$ and $6$. Fits, over the range $Q^2 < 0.40$, are done separately for each $m_f$.  }
\end{figure}

We calculate $\Pi_{V-A}(Q^2)$ for a range
of positive (space-like) $Q^2$ values, and for each $m_f$ extrapolate to $Q^2 = 0$ to determine the slope $ 4 \pi \Pi^{\prime}_{V-A}(0)$
entering the $S$ parameter. In Fig.~\ref{fig:V-A}, we show the simulation data for $\Pi_{V-A}(Q^2)$, along with fit curves. The data itself indicates that for $N_f = 2$, $\Pi^{\prime}_{V-A}(0)$ increases at smaller $m_f$ values, while for $N_f = 6$, it decreases, already suggesting a relative decrease in $S$ per electroweak doublet at $N_f = 6$. We fit the $\Pi_{V-A}(Q^2)$ data for $Q^2 < 0.4$ using a four-parameter, Pade(1,2) form (linear numerator, quadratic denominator). These fits, behaving like $1/Q^2$ at large positive $Q^2$, are shown with statistical error bands in Fig.~\ref{fig:V-A}. Each has two poles at real, negative $Q^2$, but they represent a time-like structure with cuts
 and multiple poles.
 Each fit leads to a value of  $\Pi^{\prime}_{V-A}(0)$ stable as the number of $Q^2$ points is varied.

\begin{figure}
  \includegraphics[width=85mm]{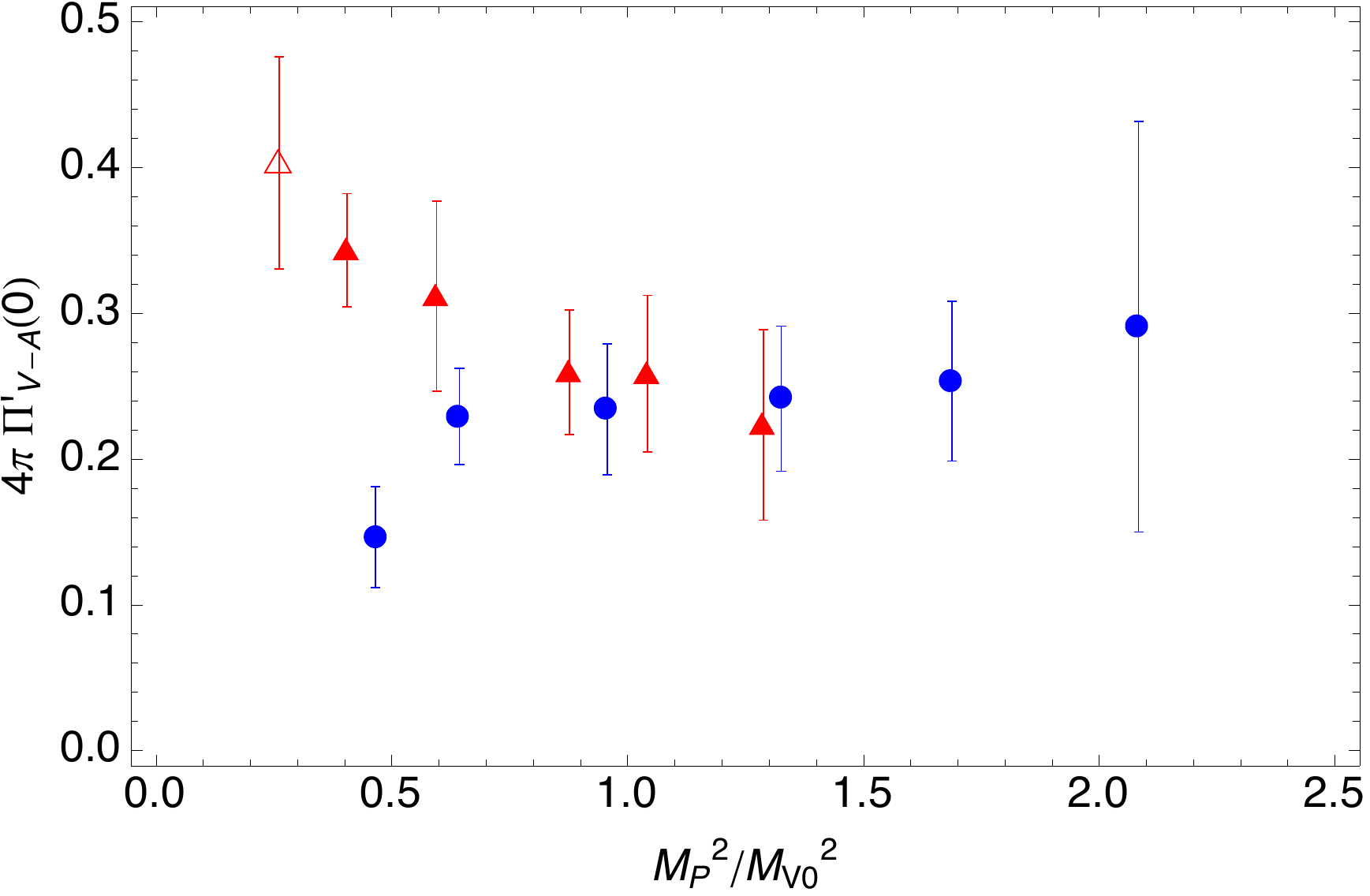}
  \caption{\label{fig:slope_vs_m}$V-A$ correlator slopes at $Q^2 = 0$ for $N_f = 2$ (red diamonds) and $N_f = 6$ (blue circles). For each of the solid points, $M_{P}L >4$.}
\end{figure}

The correlator slopes at $Q^2 = 0$ are plotted in Fig. 2. In this figure and others to follow, we plot versus $M_P^2/M_{V0}^2$ rather than $m$, where $M_P$ is the Goldstone-boson mass \cite{Appelquist:2009ka}, and $M_{V0}$ is the extrapolated mass of the lightest vector state.  We plot in this way since the relation between $M_P^2$ and $m$ is strongly $N_f$-dependent. The value of $M_{V0}$, to be discussed later, is roughly $0.2$ in lattice units for both $N_f = 2$ and $6$. For each $N_f = 6$ point and for the five heaviest $N_f = 2$ points,
$M_{P}L >4$, keeping the pion Compton wavelength well inside the lattice.

As anticipated from inspection of the data in Fig. ~\ref{fig:V-A}, $\Pi^{\prime}_{V-A}(0)$ at $N_f=6$ drops below $\Pi^{\prime}_{V-A}(0)$ at $N_f=2$ for the smaller $M_P^2$ values, suggesting a suppression of $S$ at $N_f = 6$. This interpretation requires care, however, since the extrapolation $M_P^2 \propto m \rightarrow 0$ is dominated by chiral logs ($\sim$ log$(1/m)$) for both $N_f = 2$ and $6$.

\paragraph{\textbf{S-Parameter Results}}

The $S$ parameter (Eq. \ref{eq:Sdef}) is simply the correlator slope multiplied by the number of electroweak doublets, with the SM subtraction. We estimate the SM subtraction by evaluating the $\Delta S_{SM}$ integral in Eq.~\ref{eq:Sdef} with an infrared cutoff at $s = 4M_P^2$, and taking $m_H = M_{V0}$. For the case $2M_P < M_{V0}$,
\beq\label{eq:SMsub}
\Delta S_{SM}(M_P) = \frac{1}{12\pi} \left[ \frac{11}{6} + \log \left(\frac{M_{V0}^2}{4M_P^2} \right) \right].
\eeq
We use values for $M_{P}$ and $M_{V0}$ determined in Ref. \cite{Appelquist:2009ka}.
The choice $m_H = M_{V0}$
corresponds roughly to a $1$ TeV value for the reference Higgs mass.

\begin{figure}
  \includegraphics[width=85mm]{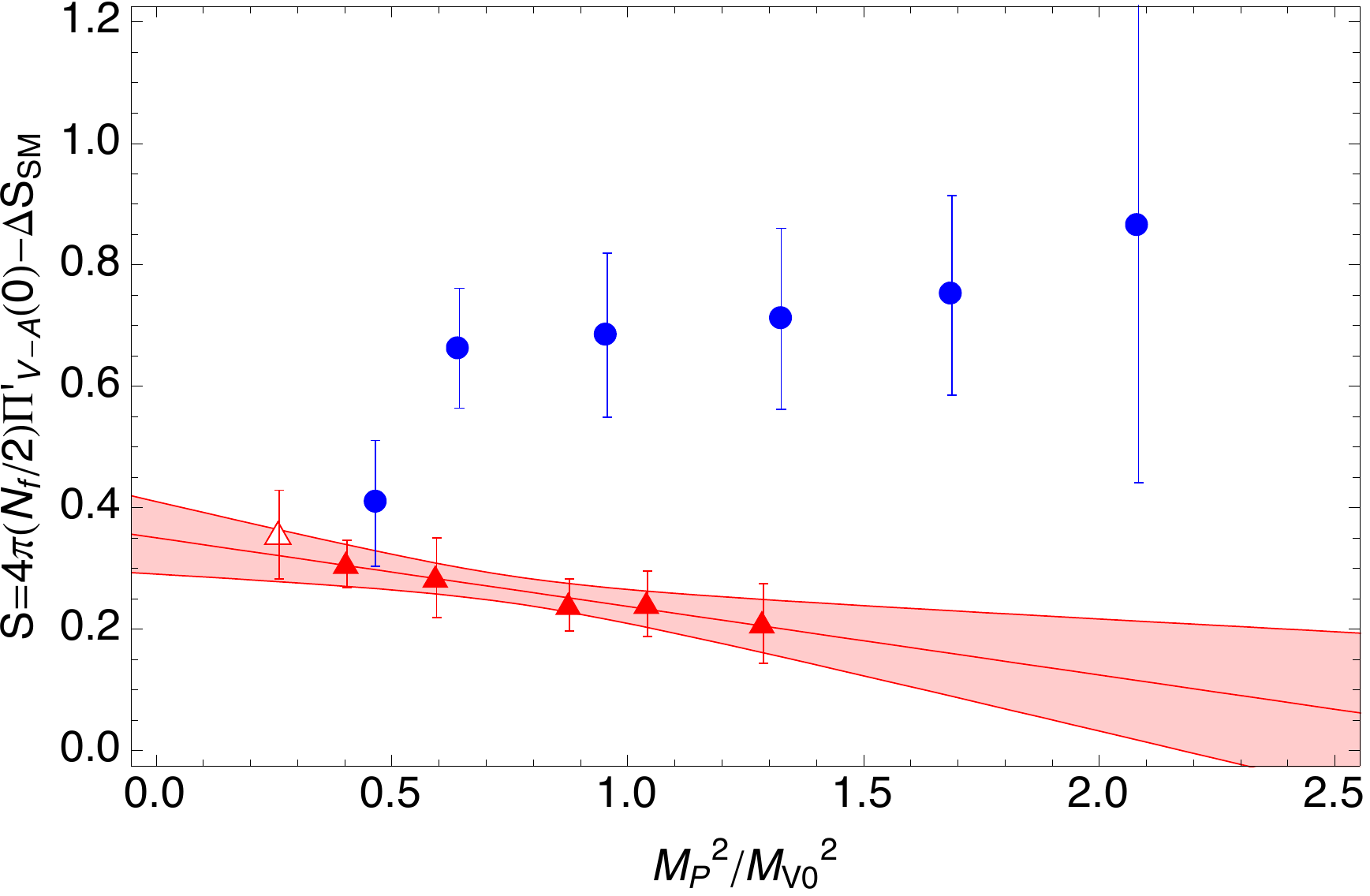}
  \caption{\label{fig:Sversusm}$S$ parameter for $N_f = 2$ (red diamonds) and $N_f = 6$ (blue circles). For each of the solid points, $M_{P}L >4$.}
\end{figure}

In Fig.~\ref{fig:Sversusm}, we plot $S \equiv 4\pi (N_f/2) \Pi'_{V-A}(0) - \Delta S_{SM}$. For $N_f = 2$, the results are consistent with previous lattice simulations \cite{Shintani:2008qe,Boyle:2009xi}. The SM subtraction at $N_f = 2$ is small, reaching a value $\sim 0.04$ for the lowest solid mass point, corresponding to $m_f = 0.010$. A smooth extrapolation to $m = 0$ is expected since the LO chiral logs eventually appearing in $\Pi'_{V-A}(0)$ are canceled by the SM subtraction, Eq. \ref{eq:SMsub}. Given the linearity and small slope of the solid data points, we include a linear fit and extrapolation. An NLO term of the form $M_P^2$log$M_P^2$ has not been ruled out, but it is not visible in our data. The fit, with error band, is shown in Fig.~\ref{fig:Sversusm}, giving $S_{m=0} = 0.35(6)$, consistent with the value obtained using scaled-up QCD data \cite{Peskin:1991sw}.

The $N_f = 6$ results for $S$ are also shown in Fig.~\ref{fig:Sversusm}. The SM subtraction is again very small as at $N_f = 2$. The important feature is that the value of $S$ at the lower mass points drops below a value obtained by simply multiplying the $N_f = 2$ result by a factor  of $3$. (For an $N_f = 6$ theory with only a single electroweak doublet, the value of $S$ at the lower $m_f$ values of Fig.~\ref{fig:Sversusm} would be well below that of the $N_f = 2$ theory.) This trend has set in at $N_f = 6$ even though $6 \ll N_{f}^c$. As $m$ is decreased further at $N_f = 6$, $S$ as computed here will eventually turn up since the SM subtraction leaves a chiral-log contribution. For $N_{f}/2$ electroweak doublets, $ S \sim (1/12\pi) [ N_f^2/4 -1]$ $\log M_P^2$. In a realistic context, the PNGBs receive mass from SM and other interactions not included here, and these masses provide the infrared cutoff in the logs.

\paragraph{\textbf{Vector and Axial Masses}}

A question of general interest for an $SU(N)$ gauge theory is the form of the resonance spectrum as $N_f$ is increased toward $N_{f}^c$. A trend toward parity doubling, for example, would provide a striking contrast with a QCD-like theory. If the gauge theory plays a role in electroweak symmetry breaking, then this trend could be associated with a diminished $S$ parameter.

\begin{figure}
  \includegraphics[width=85mm]{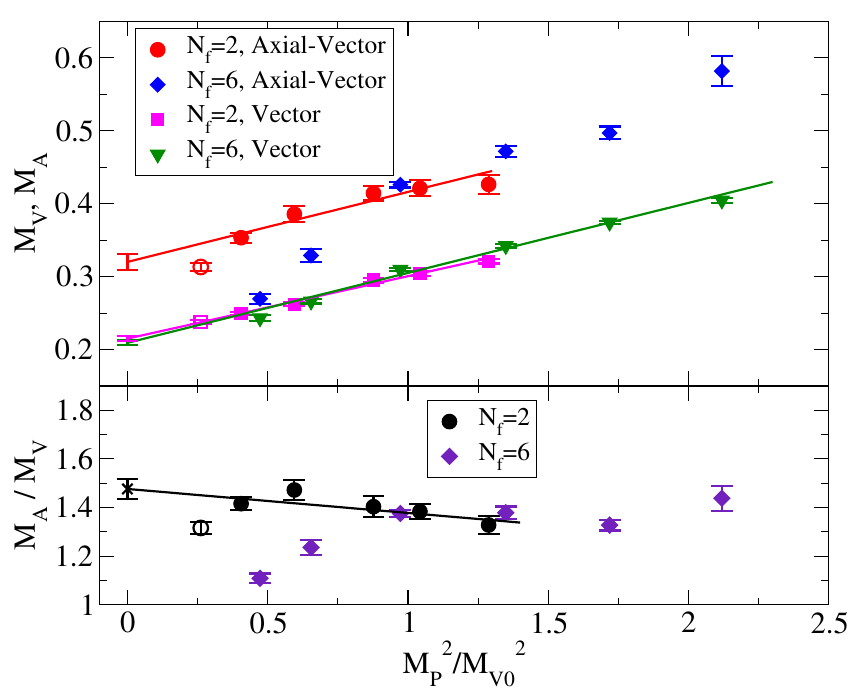}
  \caption{\label{fig:a1_rho_all}Axial and vector masses, $M_{A}$ and $M_{V}$, and their ratio. Straight
lines show linear fits to the solid points ($M_{P}L > 4$), with the
extrapolated values and errors shown to the left.}
\end{figure}

We have so far computed the masses,  $M_V$ and $M_A$, and decay constants, $F_V$ and $F_A$, of the lowest-lying vector and axial resonances. We plot the masses along with their ratio in
Fig.~\ref{fig:a1_rho_all}. Since the data points for each case except $M_{A}$ at $N_f=6$ are quite linear, with a small slope, and since in each case, the NLO term in chiral perturbation theory is linear in $M_P^2 \propto m$, we include a linear fit to the solid points ($M_{P}L > 4$). The error bars on the extrapolations are also shown. $M_V$ extrapolates to $0.215(3)$ for $N_f = 2$, and to $0.209(3)$ for $N_f = 6$.

For $N_f=2$, the extrapolated value of $M_{A}/M_V = 1.476(40)$ is
roughly consistent with the experimental result of
1.585(52)~\cite{Amsler:2008zzb}.
The $N_f = 6$ data points for $M_A$ do not yet allow a simple fit and extrapolation, but they
do indicate a substantial decrease in $M_{A}/M_{V}$ in the chiral limit. This trend toward parity doubling
suggests that the spectrum could become even more parity doubled as $N_f$ is increased further, toward $N_{f}^c$.

\paragraph{\textbf{Vector and Axial Decay Constants}}

\begin{figure}
  \includegraphics[width=85mm]{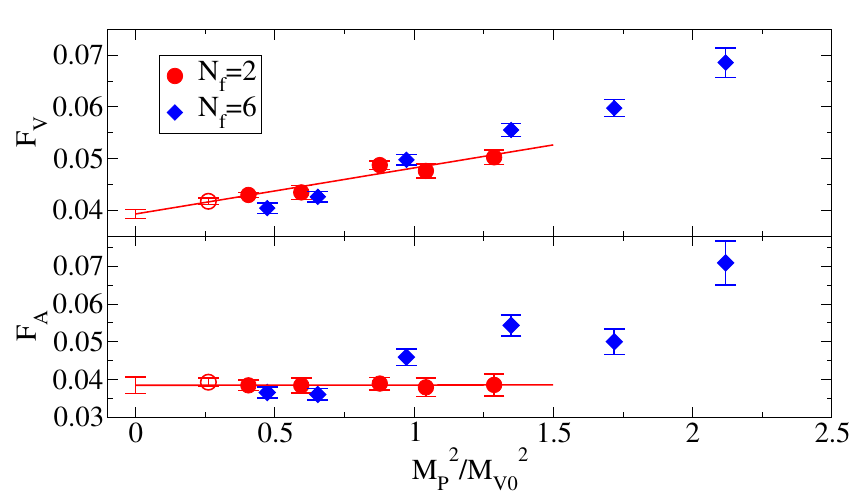}
  \caption{\label{fig:Fa1_Frho}Decay constants $F_{V}$ and $F_{A}$.
  The straight lines represent linear fits to the $N_f=2$
data, with the extrapolated values in the chiral limit shown at the
left end of the lines.}
\end{figure}

Our simulation results for $F_V$ and $F_{A}$ are shown in Fig. ~\ref{fig:Fa1_Frho},
using the normalization conventions of Ref. \cite{Peskin:1991sw}. The dependence on
$M_P^2/M_{V0}^2$ is mild, and once again, for each case except the $A$ at $N_f = 6$, quite linear with a small slope. Although there is known to be an NLO chiral log for the decay constants, it is not visible in the linear points, so we include a linear fit to the $N_f = 2$ data. The linearly
extrapolated values, converted to physical units using the
lattice scale determined from
$M_{V0}$, are $F_V = 141.8(3.8)$
MeV and $F_{A} = 138.9(8.2)$ MeV, agreeing well with the measured QCD results \cite{Allton:2008pn,Isgur:1988vm}.

\paragraph{\textbf{Discussion}}

The relation between a diminished $S$ parameter and the spectrum can be explored through the dispersion relation
\begin{equation}
 \Pi_{V-A}(Q^2) = \frac{Q^2}{12\pi} \int_0^\infty \frac{ds}{\pi} \frac{ R_V(s) - R_A(s)}{s+Q^2} - F_{P}^2 \,, \label{eq:Spectral}
\end{equation}
where $F_{P} \simeq 93$ MeV for two-flavor QCD.
Here, we include a few remarks about this use of the dispersion relation
via the assumption of single-pole dominance

As a preliminary, recall that in the chiral limit (and in the continuum), the operator product expansion gives $\Pi_{V-A}(Q^2) \sim 1/Q^4$ as $Q^2 \rightarrow \infty$. In the resultant first and second Weinberg sum rules (WSRs), the assumption of single pole dominance, $R_{V,A}(s) = 12 \pi^2 F_{V,A}^{2}\delta(s - M_{V,A}^2)$, then leads to the relations $F_{V}^2 - F_{A}^2 = F_{P}^2$ and $F_{V}^{2}M_{V}^2- F_{A}^{2} M_{A}^2= 0$. At finite $m$, however, $\Pi_{V-A}(Q^2)$ falls more slowly with $Q^2$. The simulation data in the $Q^2$ range of Fig. \ref{fig:V-A} behaves like $m \langle {\bar \psi} \psi \rangle /Q^2$, but there will be an additional small ($O(m^2)$) term \cite{Shifman:1978by}. While the data are consistent with the first WSR, the single-pole-dominance relation $F_{V}^2 - F_{A}^2 = F_{P}^2$ is not satisfied at $N_f = 2$ for any $m$, the LHS being less than $50\%$ of the RHS. The extrapolated values and the QCD experimental values fail by even more. A similar discussion applies to the second WSR at $N_f = 2$ and the resultant pole-dominance relation. For $N_f = 6$, the errors on $F_{A}$ do not yet allow a useful test.


 A crude expression for $S$, which derives from the $Q^2 \rightarrow 0$ limit of the dispersion relation, can be obtained by assuming single-pole dominance, and neglecting
 the SM subtraction. The result, $S = 4 \pi [ F_{V}^{2}/M_{V}^2- F_{A}^{2}/ M_{A}^2 ]$, pays no heed to a reference Higgs mass and has no direct dependence on the PNGB masses for $N_f = 6$. On the other hand, the stronger UV convergence of the integral expression (Eq. \ref{eq:Sdef}) could make single-pole dominance more reliable than for the WSRs. In fact, evaluation of the above expression using our simulation data leads, for $N_f=2$, to a result within $30\%$ of our direct simulation of $S$ at the smaller $M_P^2$ values. For $N_f = 6$, the agreement with the direct simulation is at least as good, and also shows the relative decrease of $S$ per electroweak doublet.

\paragraph{\textbf{Conclusions}}

We have described a lattice simulation of the masses and decay constants of the lowest-lying vector and axial resonances, and the electroweak $S$ parameter, in an $SU(3)$ gauge theory with $N_f = 2$ and $6$ fermions in the fundamental representation. The spectrum becomes more parity doubled and the $S$ parameter per electroweak doublet decreases when $N_f$ is increased from $2$ to $6$. The final value of $S$ for any $N_f >2$ will depend logarithmically on the masses of PNGBs, generated by SM and other physics not included here.
It will be interesting to study these trends as $N_f$ is increased further, toward $N_{f}^c$.

\paragraph{\textbf{Acknowledgements}}We thank the Aspen Center
 for Physics, and the LLNL Multiprogrammatic and Institutional Computing
 program for time on the BlueGene/L supercomputer.  This work was supported by the NNSA and Office of Science of the U.S. Department of
 Energy, and by the U.S. National Science Foundation.

\bibliography{LSD-S-param}


\end{document}